\documentclass[sigconf,screen]{acmart}
\AtBeginDocument{%
  }

\setcopyright{cc}
\setcctype{by}
\acmDOI{10.1145/3805760.3814923}
\acmYear{2026}
\copyrightyear{2026}
\acmISBN{979-8-4007-2601-9/2026/07}
\acmConference[AIware '26]{Proceedings of the 3rd ACM International Conference on AI-Powered Software}{July 6--7, 2026}{Montreal, QC, Canada}
\acmBooktitle{Proceedings of the 3rd ACM International Conference on AI-Powered Software (AIware '26), July 6--7, 2026, Montreal, QC, Canada}
\acmSubmissionID{fseaiware26main-pp027-data-p}
\received{2026-02-15}
\received[accepted]{2026-03-28}

\newcommand{\nd}{\vspace{1mm}\noindent}

\usepackage{enumitem}
\usepackage{multirow}
\usepackage{framed}
\usepackage[ruled,vlined,linesnumbered]{algorithm2e}
\usepackage{minted}
\usepackage{fontawesome}
\usepackage{listings}

\definecolor{summarygray}{gray}{0.90}

\newenvironment{resultbox}
  {%
   \MakeFramed{\advance\hsize-\width \FrameRestore}}
  {\endMakeFramed}

\begin{document}

\title[\textsc{AgenticFlict}: A Large-Scale Dataset of Merge Conflicts in AI Coding Agent Pull Requests on GitHub]{\textsc{AgenticFlict}: A Large-Scale Dataset of Merge Conflicts in AI Coding Agent Pull Requests on GitHub}

\author{Daniel Ogenrwot}
\orcid{0000-0002-0133-8164}
\affiliation{
  \institution{University of Nevada Las Vegas}
  \city{Las Vegas}
  \country{USA}
}
\email{ogenrwot@unlv.nevada.edu}

\author{John Businge}
\orcid{0000-0003-3206-7085}
\affiliation{
  \institution{University of Nevada Las Vegas}
  \city{Las Vegas}
  \country{USA}
}
\email{john.businge@unlv.edu}

\renewcommand{\shortauthors}{Daniel Ogenrwot and John Businge}

\begin{abstract}
Software Engineering 3.0 marks a paradigm shift in software development, in which AI coding agents are no longer just assistive tools but active contributors. While prior empirical studies have examined productivity gains and acceptance patterns in AI-assisted development, the challenges associated with integrating agent-generated contributions remain less understood. In particular, \textbf{merge conflicts}, a fundamental aspect of collaborative software development, remain underexplored in this context. In this paper, we present \textsc{AgenticFlict}, a large-scale dataset of textual merge conflicts in AI coding agent pull requests (Agentic PRs). The dataset comprises 142K+ Agentic PRs collected from 59K+ repositories, of which 107K+ are successfully processed through deterministic merge simulation. Our pipeline identifies 29K+ PRs exhibiting merge conflicts, yielding a conflict rate of 27.67\%, and extracts 336K+ fine-grained conflict regions across these instances. Our preliminary exploratory analysis indicates that merge conflicts are both frequent and often substantial in AI-generated contributions, with noticeable variation across agents, emphasizing the need to better understand and manage integration challenges in AI-assisted software development. \textbf{The dataset, code and supplementary materials are available in zenodo:~\href{https://doi.org/10.5281/zenodo.19396916}{10.5281/zenodo.19396916}}
\end{abstract}



\maketitle

\section{Introduction}
The rise of Artificial intelligence (AI) coding agents is reshaping modern software development workflows. Several AI coding tools such as GitHub Copilot~\cite{githubCopilot}, OpenAI Codex~\cite{openaiCodexWeb}, Claude Code~\cite{claudeAI2025}, Cursor~\cite{cursor2025}, and Devin~\cite{devinAIWeb} assist developers by generating code, suggesting refactorings, and increasingly contributing changes in the form of pull requests (PRs). This evolution reflects a broader shift from assistive tooling toward active collaboration, often described as Software Engineering 3.0~\cite{hassan2025agenticsoftwareengineeringfoundational,hassan2024ainativesoftwareengineeringse,ogenrwot2026aicodingagentsmodify}. Prior work has examined how developers interact with AI-generated code and the impact of these tools on productivity and software quality~\cite{vaithilingam2022expectation,li2025aidev,vaithilingam2023copilot,ogenrwot2025patchtrackcomprehensiveanalysischatgpts, ogenrwot:2024ase}. These studies highlight both the opportunities and challenges associated with human–AI collaboration. More recent empirical research has begun to investigate development efficiency and code review dynamics in AI-assisted settings~\cite{Vijayvergiya2024AI,ogenrwot2025patchtrackcomprehensiveanalysischatgpts,10.1145/3520312.3534864}. However, a fundamental aspect of collaborative software engineering, namely \textbf{merge conflicts}, remains largely unexplored in the context of AI-generated contributions.

In practice, integrating code in a collaborative environment is rarely smooth due to conflicts~\cite{Shen2023A,ogenrwot:2025:scam,businge:emse:2022,Businge:benevol:2020,Businge:chap:2023}. Merge conflicts arise when concurrent modifications affect overlapping regions of code and cannot be automatically reconciled by version control systems like Git. Prior research has shown that merge conflicts introduce substantial coordination overhead and negatively impact developer productivity~\cite{brun2013early,6227180,9625780,8094445,8870173}. Brun et al.~\cite{brun2013early} demonstrate that collaboration conflicts are frequent and costly in distributed development environments. Gousios et al.~\cite{Gousios:2014} analyze pull-based development workflows and highlight the complexity of integrating contributions through PRs. Studies on modern code review further emphasize that integration friction influences review latency and decision making~\cite{kononenko2016code,watanabe2025useagenticcodingempirical}.

Despite the growth of empirical research in this area, curated datasets specifically targeting merge conflicts remain limited. While \citet{SHEN2024112084} introduced ConflictBench as a dedicated benchmark for studying conflicts, other existing datasets have typically emerged as secondary artifacts of broader studies on collaborative development~\cite{8468085,10.1145/3540250.3549163,10.1145/3555228.3555229}. More recently, \citet{li2025aidev} presented AIDev, a large-scale dataset capturing PRs (\textit{a.k.a} Agentic PRs), issues, and discussions involving five AI coding agents. However, these datasets fail to provide explicit, reproducible labels for textual merge conflicts; instead, they prioritize metrics such as acceptance rates, temporal dynamics, and general repository characteristics. \citet{watanabe2025useagenticcodingempirical} report that merge conflicts account for over 1.1\% of Agentic-PR rejections. A concrete example is observed in~\href{https://github.com/openai/codex/pull/612}{\faGithub~openai/codex-PR\#612}
, where the pull request was abandoned because the contributor was unable to resolve the merge conflict.

Researchers currently lack the necessary resources to study integration friction introduced by AI coding agents in collaborative software engineering environments. To address this gap, we introduce \textsc{AgenticFlict}, a large-scale dataset of textual merge conflicts in AI coding agent PRs derived from AIDev dataset~\cite{li2025aidev}. The dataset comprises 142{,}652 Agentic PRs collected from 59{,}412 repositories, of which 107{,}026 are successfully processed through deterministic merge simulation of open and/or closed (unmerged) PRs. Our pipeline identifies 29{,}609 PRs exhibiting merge conflicts, yielding a conflict rate of 27.67\%, and extracts 336{,}380 fine-grained conflict regions across these instances. Beyond binary conflict labels, \textsc{AgenticFlict} provides detailed conflict-region metadata, including affected file paths, conflict regions, and line-level spans. 
The dataset spans contributions from five distinct AI coding agents, enabling comparative analysis of conflict prevalence and severity across agents. 

To the best of our knowledge, \textsc{AgenticFlict} is the first large-scale dataset of textual merge conflicts in Agentic PRs. This dataset can support several research directions, including: (i) empirical studies of merge conflict prevalence and characteristics in AI-generated code; (ii) comparative analysis of integration behavior across AI coding agents; (iii) training and evaluation of automated conflict detection and resolution models; (iv) analysis of the relationship between pull request characteristics (e.g., size, files changed) and conflict likelihood or severity; and (v) benchmarking tools for conflict prediction, merge automation, and collaborative development support in AI-assisted workflows.

In summary, the \textbf{contributions} of this work are as follows:

\begin{itemize}[leftmargin=*]
    \item A reproducible merge simulation pipeline for large-scale conflict detection in pull requests.
    \item A pull request level dataset containing textual conflict labels and severity metrics.
    \item A fine-grained conflict-region dataset with file paths and exact line spans of conflicting regions.
    \item A publicly released artifact to support research on AI-assisted collaboration and integration friction.
\end{itemize}

\section{Dataset Curation Methodology}
\label{sec:dataset_construction}

\begin{figure}
    \centering
    \includegraphics[width=\linewidth]{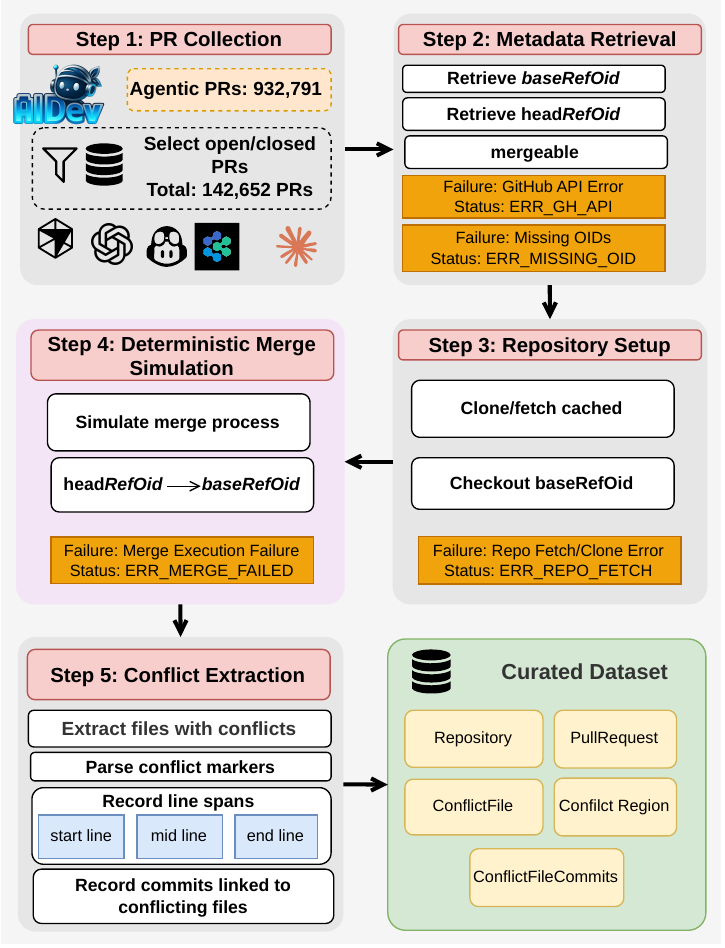}
    \caption{Overview of the \textsc{AgenticFlict} dataset curation workflow.}
    \Description{Pipeline diagram illustrating ingestion of AIDev PR records, metadata retrieval from GitHub, repository setup, merge simulation and conflict extraction.}
    \label{fig:data-curation}
\end{figure}

Figure~\ref{fig:data-curation} summarizes the multi-stage workflow used to construct the \textsc{AgenticFlict} dataset. 
The pipeline consists of five main stages:  (1) Agentic PR collection from the AIDev dataset, (2) Metadata Retrieval, (3) repository preparation, (4) deterministic merge simulation, and (5) conflict extraction.

\nd \textbf{Step 1: Pull Request Collection.}
We use the AIDev dataset~\cite{li2025aidev} downloaded from \href{https://huggingface.co/datasets/hao-li/AIDev}{\faGlobe \hspace{0.01in} Hugging Face} as of January 5, 2026. The dataset contains 932{,}791 Agentic PRs. As an initial preprocessing step, we retain PRs that are either open or closed without evidence of having been merged. When raw state and merge timestamps are available, this filtering is performed before the extraction pipeline begins; otherwise, the final decision is deferred to GitHub metadata retrieval in next step of the pipeline. This filtering yielded \textbf{142,652} candidate PRs. Although this design may slightly reduce dataset coverage, it guarantees that all retained records correspond to verifiable GitHub artifacts. Each pull request is identified by a repository name (\texttt{repo\_full\_name}), in the form (\texttt{owner\//repository}) and a pull request number (\texttt{pr\_number}). We combine these to construct a canonical identifier (\texttt{pr\_key}) of the form \texttt{repo\_full\_name\#pr\_number}, enabling consistent tracking throughout the pipeline.

\nd \textbf{Step 2: Metadata Retrieval.}
For each pull request, we query the GitHub GraphQL API~\cite{github-graphql-doc} to retrieve repository and pull request metadata, including the pull request state, timestamps, branch names, and the base and head commit object identifiers (\texttt{baseRefOid} and \texttt{headRefOid}), which serve as the primary anchors for merge simulation. At scale, interacting with the GitHub API introduces several practical limitations. In particular, requests may fail due to rate limiting (HTTP 403)~\cite{github-rate-limit,7887704,10.1145/2597073.2597074}, transient server errors (e.g., HTTP 502/503), or repository-level issues such as deletion or restricted access (HTTP 404/410/451)~\cite{10.1145/2597073.2597074,7887704}. To mitigate these challenges, our implementation employs bounded retries with increasing delays to mitigate transient API failures, token rotation to distribute request load, and explicit handling of API error codes. Despite these safeguards, some PRs remain unrecoverable due to permanently missing references or inaccessible repositories. We identified 35,626 such cases. Instead of silently discarding them, we explicitly record failure modes using structured status codes and exclude these instances from downstream conflict analysis.

\nd \textbf{Step 3: Repository Preparation.}
Before merge simulation, each repository is prepared locally. We clone repositories into a persistent cache using Git's partial clone mechanism~\cite{git-partial-clone}, which downloads repository history while avoiding unnecessary file blobs. 
Subsequent PRs belonging to the same repository reuse the cached clone and perform a lightweight \texttt{git fetch} to synchronize the repository state.

For each pull request, the pipeline resets the working tree to a clean state and checks out the base commit identified by \texttt{baseRefOid}. 
This preparation step ensures that every merge simulation begins from a deterministic repository state and avoids interference from previous operations.

\nd \textbf{Step 4: Deterministic Merge Simulation.}
Algorithm~\ref{alg:merge_reconstruction} summarizes this step. To determine whether a pull request produces a textual merge conflict, we perform a local merge simulation using Git. Given the base and head commit OIDs retrieved from GitHub, we execute the following command: \texttt{git merge --no-commit --no-ff <headRefOid>}. If the merge completes successfully, the pull request is labeled \texttt{merge\_clean}. If the merge fails, the repository enters a conflicted state and we proceed to conflict extraction. 

The simulation procedure differs slightly depending on pull request state. 
For open PRs, we simulate the merge using the current base and head commit OIDs returned by the API. 
For closed but unmerged PRs, the base branch may have advanced after closure; therefore, we reconstruct the base commit corresponding to the repository state at the time the pull request was closed and perform the merge against that snapshot. 
This design approximates the merge conditions that developers would have encountered at closure time.

Certain merge simulation failures may arise when repositories have been deleted or historical commits are no longer reachable due to force-pushes or history rewrites~\cite{5069475,businge:2018icsme,businge:blockchain:2022}. Such cases are explicitly labeled with structured error codes and recorded in the dataset's run log, allowing downstream analyses to quantify merge simulation coverage.








\begin{algorithm}[t]
\caption{Deterministic Merge Simulation and Conflict Extraction}
\label{alg:merge_reconstruction}
\KwIn{Repository path $R$, base commit $b$, head commit $h$}
\KwOut{Merge outcome, conflict metrics, and extracted conflict regions}
Reset the working state of repository $R$\;
Checkout base commit $b$\;
Create a temporary analysis branch\;
$rc \gets$ \textsc{SimulateMerge}($R, h$)\;
\If{$rc = \textsc{Success}$}{
    Revert temporary merge state\;
    \Return \texttt{merge\_clean}, empty metrics, empty region set\;
}
$F \gets$ \textsc{ListConflictedFiles}($R$)\;
$Regions \gets \emptyset$\;
Initialize metrics:\;
\quad $num\_conflict\_files \gets 0$\;
\quad $num\_conflict\_regions \gets 0$\;
\quad $conflict\_lines \gets 0$\;
\ForEach{$f \in F$}{
    $text \gets$ \textsc{ReadFile}($R, f$)\;
    $R_f \gets$ \textsc{ParseConflictRegions}($text$)\;
    Add $R_f$ to $Regions$\;
    Update metrics using the extracted regions from $R_f$\;
}
Revert temporary merge state\;
\Return \texttt{merge\_conflict}, metrics, $Regions$\;
\end{algorithm}

\nd \textbf{Step 5: Conflict Detection and Region Extraction.}
When a merge operation fails, Git records unresolved files in the index. As indicated in Line 9 of the Algorithm~\ref{alg:merge_reconstruction}, 
we identify these files using: \texttt{git diff --name-only --diff-filter=U}.
Each conflicted file contains standard Git conflict markers:
\verb|<<<<<<<|, \verb|=======|, and \verb|>>>>>>>|. 
We parse these markers to extract structured conflict regions. For each region, we record several parameters including: file path, conflict index within the file, line boundaries (\texttt{start\_line}, \texttt{mid\_line}, \texttt{end\_line}), SHA-256 hashes of each side's code block, and short textual previews of each side. In addition, we also compute PR-level severity metrics such as the number of conflicting files, number of conflict regions, and total number of lines contained within conflict markers.

To balance dataset size and comply with repository licensing constraints, we store compact representations of conflicts, including content hashes and short previews (default: 5 lines of code), rather than full conflict blocks. This approach follows established practices in large-scale mining of GitHub data~\cite{di2017software,GHTorrent,Svajlenko:2014:ICSME} and aligns with GitHub's Terms of Service governing code redistribution~\cite{github-tos}.

Beyond identifying conflict regions, we attribute each conflicting file to the most recent commit that modified the file on both the base and head sides of the merge. 
This attribution is computed using: \texttt{git log -n 1 --format=\%H <rev> -- <file>}. The resulting fields \texttt{head\_last\_touch\_oid} and \texttt{base\_last\_touch\_oid} provide a lightweight proxy for identifying the commits most directly associated with the conflicting file. Although this approach does not perform line-level blame alignment, it provides sufficient granularity for studying commit structuring, change locality, and conflict concentration.

\section{Dataset Schema and Dataset Overview}
\label{sec:dataset_schema}

\begin{figure*}[ht]
    \centering
    \includegraphics[width=\linewidth]{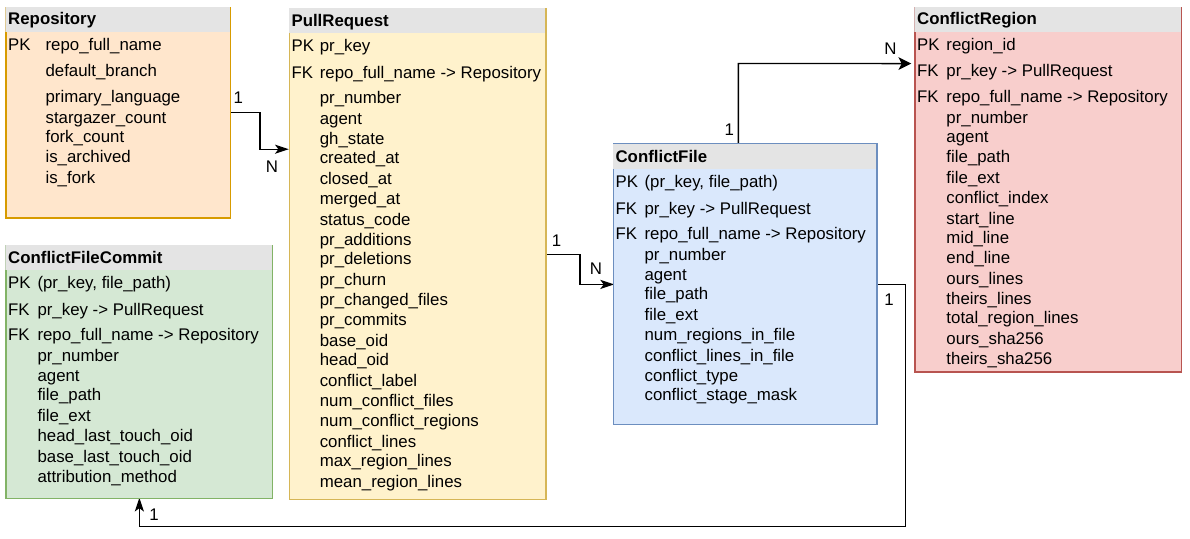}
    \caption{Dataset Schema of \textsc{AgenticFlict}.}
    \Description{Dataset Schema of \textsc{AgenticFlict}.}
    \label{fig:ERD}
\end{figure*}
 The dataset is organized as a relational schema supporting analysis at multiple levels of granularity. We provide both a raw dataset, which includes full pipeline metadata, and a clean dataset, which retains only analysis-relevant attributes. The discussions and results in this paper are based on the clean dataset. A detailed mapping of retained and removed fields is included in the replication package~\cite{AgenticFlict2026Replication}.

\subsection{Schema Overview}

We describe the schema of \textsc{AgenticFlict}, illustrated in Figure~\ref{fig:ERD}. The schema consists of five primary entities, which are explained below. Additional details on field definitions are provided in the replication package as an online appendix~\cite{AgenticFlict2026Replication}.

\nd \textbf{Repository.}  
The repository entity stores contextual metadata about repositories referenced in the dataset, including repository name, star count, fork count, primary programming language, and repository status (e.g., archived or fork). Separating this information avoids redundancy when multiple PRs originate from the same repository.

\nd \textbf{PullRequest.}  
The PR entity is the central component of the dataset and contains one record per pull request. It stores GitHub metadata such as repository identifier, pull request number, state, timestamps, and mergeability signals. In addition, it records reconstruction outcomes, including whether a conflict occurs and aggregate severity metrics such as the number of conflicting files, number of conflict regions, and total conflict lines.

\nd \textbf{ConflictFile.}  
This entity captures file-level conflict information and is linked to the pull request entity via \texttt{pr\_key}. Each record corresponds to a file containing at least one conflict region and includes attributes such as the number of conflict regions, total conflict lines, file extension, and conflict type (e.g., both-modified, modify/delete).

\nd \textbf{ConflictRegion.}  
Provides fine-grained conflict details. Each record represents a single conflict region within a file and includes the file path, region index, and line-level boundaries (\texttt{start\_line}, \texttt{mid\_line}, \texttt{end\_line}). Additional attributes capture the size of each side of the conflict and compact hash representations of the conflicting code blocks.

\nd \textbf{ConflictFileCommit.}  
This entity links conflicting files to the commits most recently modifying them on each side of the merge. For each conflicting file, we record the last commit touching the file on the head and base branches. This provides a lightweight approximation of the origins of conflicting changes and enables analyses of conflict provenance.

\subsection{Dataset Overview}
Table~\ref{tab:dataset_stats} provides an overview of the \textsc{AgenticFlict} dataset. Starting from 142,652 Agentic PRS, we successfully performed merge simulation for 107,026 instances, corresponding to a success rate of 75.03\%. The remaining PRs were excluded due to missing commit references or repository access limitations, which are common challenges when working with large-scale GitHub data~\cite{5069475,10.1145/2597073.2597074,10.1145/3540250.3549112}.

Among the successfully simulated PRs, we observe that merge conflicts are relatively frequent. In particular, 27.67\% of PRs result in textual conflicts, indicating that integration issues are not uncommon in AI-generated contributions. This suggests that, despite their usefulness, AI coding agents can introduce non-trivial challenges during code integration.

We further examine the severity of these conflicts by focusing on PRs that exhibit conflicts. On average, a conflicting pull request affects 4.36 files, with a median of 2 files, indicating that most conflicts are relatively localized, but a subset involves multiple files. Each conflicting pull request contains an average of 11.36 conflict regions and over 500 conflicting lines, suggesting that conflicts are often substantial rather than isolated. Overall, the dataset contains more than 336,000 fine-grained conflict regions.

Finally, the dataset spans 59,412 distinct repositories and includes contributions from five different AI coding agents. This diversity provides a broad view of how Agentic PRS behave across different projects and development contexts, supporting comparative and large-scale empirical analyses.
\begin{table}[ht]
\centering
\caption{Summary statistics of the \textsc{AgenticFlict} dataset.}
\label{tab:dataset_stats}
\setlength{\tabcolsep}{6pt}
\begin{tabular}{l r}
\toprule
\textbf{Metric} & \hspace{2.5cm}\textbf{Value} \\
\midrule

\multicolumn{2}{l}{\textit{\textbf{Dataset Overview}}} \\
Total AI PRs identified & 142{,}652 \\
Valid PRs with identifiers  & 142{,}652 \\
Successfully simulated PRs  & 107{,}026 \\
Excluded PRs & 35{,}626 \\
Merge simulation success rate & 75.03\% \\

\midrule
\multicolumn{2}{l}{\textit{\textbf{Merge Outcomes}}} \\
Conflicting PRs & 29{,}609 \\
Clean PRs & 75{,}924 \\
Conflict rate & 27.67\% \\

\midrule
\multicolumn{2}{l}{\textit{\textbf{Conflict Severity (per conflicting PR)}}} \\
Mean conflicting files per PR & 4.36 \\
Median conflicting files per PR & 2.00 \\
Mean conflict regions per PR & 11.36 \\
Mean conflict lines per PR & 540.42 \\
Total conflict regions & 336{,}380 \\

\midrule
\multicolumn{2}{l}{\textit{\textbf{Dataset Diversity}}} \\
Distinct repositories & 59{,}412 \\
Distinct AI agents & 5 \\

\bottomrule
\end{tabular}
\end{table}

\section{Exploratory Empirical Analysis}

We perform an exploratory empirical analysis to characterize merge conflict behavior in Agentic PRS. Specifically, we investigate (1) how pull request size relates to the likelihood of merge conflicts, and (2) how conflict rates and severity vary across different AI coding agents. These analyses provide initial evidence on how change characteristics and agent behavior influence integration outcomes in AI-assisted software development.

\nd \faQuestionCircle \hspace{0.02in} \textbf{How do merge conflict rates and severity vary across AI Coding Agents?}
To understand whether different AI coding agents exhibit distinct integration behaviors, we analyze conflict rates and conflict severity across agents.

\begin{table*}[ht]
    \centering
     \caption{Conflict rates across AI coding agents with 95\% confidence intervals.}
    \begin{tabular}{lrrrrr}
    \toprule
    Agent & \hspace{2cm} PRs & \hspace{0.8cm} Conflicting PRs & \hspace{0.8cm} Conflict Rate (\%) & \hspace{0.8cm} 95\% CI Low & \hspace{0.8cm} 95\% CI High \\
    \midrule
    Copilot & 16954 & 2583 & 15.24 & 14.69 & 15.78 \\
    Cursor & 7196 & 1421 & 19.75 & 18.83 & 20.67 \\
    Devin & 8241 & 1883 & 22.85 & 21.94 & 23.76 \\
    Claude\_Code & 779 & 202 & 25.93 & 22.85 & 29.01 \\
    OpenAI\_Codex & 73856 & 23520 & 31.85 & 31.51 & 32.18 \\
    \bottomrule
    \end{tabular}
    \label{tab:agent_conflict_rate}
\end{table*}

Table~\ref{tab:agent_conflict_rate} summarizes the number of PRs, conflicting PRs, and corresponding conflict rates with 95\% confidence intervals for each agent. We observe substantial variation across agents. Copilot exhibits the lowest conflict rate at 15.43\%, followed by Cursor (20.06\%) and Devin (23.04\%). In contrast, OpenAI Codex shows the highest conflict rate at 32.31\%, more than double that of Copilot. Claude\_Code also demonstrates relatively high conflict rates (26.86\%), although with wider confidence intervals due to smaller sample size.

\begin{figure}[ht]
    \centering
    \includegraphics[width=\linewidth]{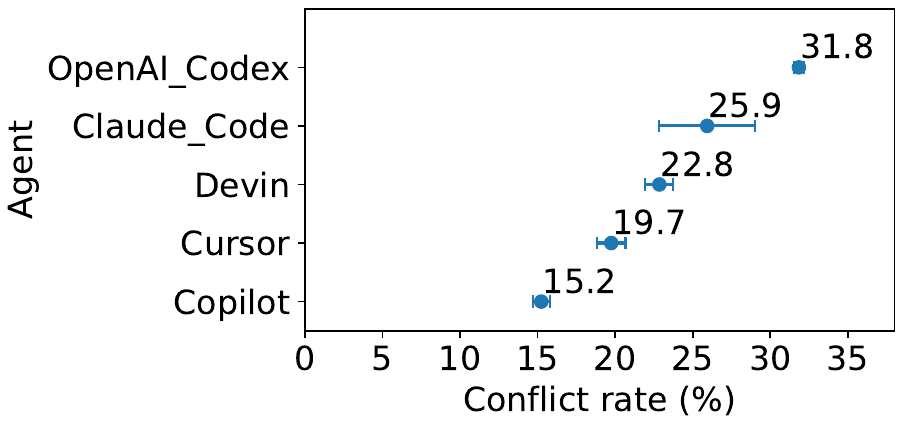}
    \caption{Conflict rates across AI coding agents with 95\% confidence intervals.}
    \Description{Conflict rates across AI coding agents with 95\% confidence intervals.}
    \label{fig:conflict_rate_by_agent}
\end{figure}

Figure~\ref{fig:conflict_rate_by_agent} visualizes these differences with confidence intervals, highlighting that the variation is statistically meaningful. The separation between agents, particularly between Copilot and OpenAI Codex, suggests that the likelihood of introducing merge conflicts varies significantly depending on the underlying AI system.

\begin{figure}[ht]
    \centering
    \includegraphics[width=\linewidth]{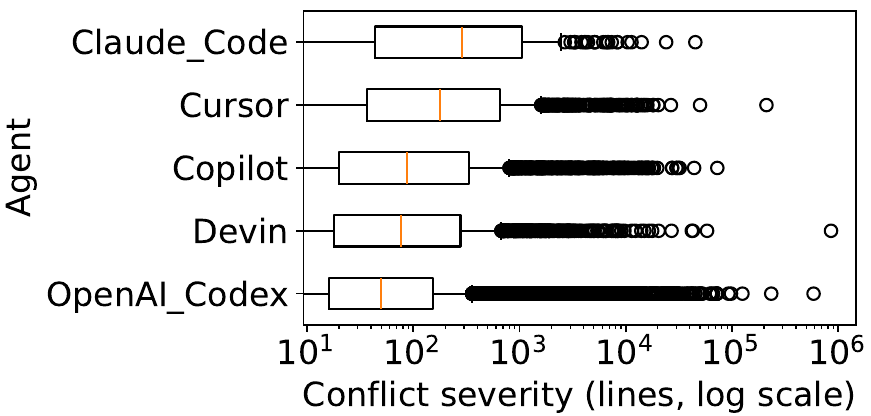}
    \caption{Distribution of conflict severity (measured as conflicting lines) across AI coding agents.}
    \Description{Distribution of conflict severity (measured as conflicting lines) across AI coding agents.}
    \label{fig:conflict_severity_by_agent}
\end{figure}

To further examine the nature of these conflicts, we analyze conflict severity, measured as the number of conflicting lines per PR. Figure~\ref{fig:conflict_severity_by_agent} shows the distribution of conflict severity across agents on a logarithmic scale. We observe heavy-tailed distributions for all agents, indicating that while most conflicts are relatively small, some PRs introduce very large and complex conflicts. Notably, OpenAI Codex exhibits both a higher conflict rate and a broader spread of conflict severity, suggesting that it not only conflicts more frequently but may also produce more complex integration challenges.

Overall, these findings indicate that AI coding agents differ not only in how often they produce conflicting changes, but also in the magnitude of those conflicts. This highlights the importance of considering agent-specific behaviors when designing integration workflows and evaluation benchmarks for AI-assisted software development.

\begin{resultbox}
\noindent\textbf{\faLightbulbO \hspace{0.01in} Key takeaway: }
AI coding agents differ in both the frequency and severity of merge conflicts, highlighting the need for agent-aware integration workflows and evaluation strategies.
\end{resultbox}

\nd \faQuestionCircle \hspace{0.02in} \textbf{How does pull request size affect merge conflict likelihood?} We investigate the relationship between PR size and the likelihood of merge conflicts. We measure PR size using \emph{code churn}, defined as the sum of lines added and deleted in a pull request. To analyze this relationship, we group PRs into deciles based on churn and compute the conflict rate within each bin.

\begin{figure}[ht]
    \centering
    \includegraphics[width=\linewidth]{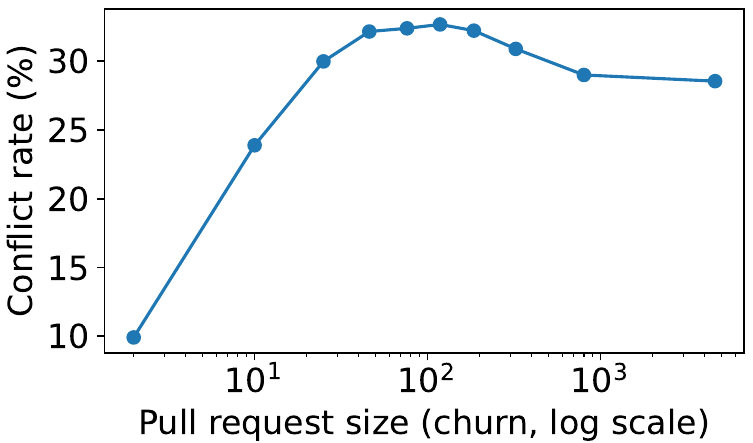}
    \caption{Conflict rate as a function of pull request size (measured as code churn). PRs are grouped into deciles based on size.}
    \label{fig:pr_size_vs_conflict}
\end{figure}

Figure~\ref{fig:pr_size_vs_conflict} shows the resulting relationship between PR size and conflict rate. We observe a clear trend: smaller PRs are significantly less likely to exhibit merge conflicts, while conflict rates increase rapidly as PR size grows. For example, PRs with a median churn of 2 lines have a conflict rate of approximately 9.9\%, whereas PRs with a median churn of 25 lines exhibit a conflict rate of nearly 30\%. 

The conflict rate continues to increase and stabilizes around 32--33\% for medium-sized PRs (median churn between 46 and 185 lines). For larger PRs, the conflict rate slightly decreases but remains substantially higher than that of small PRs, suggesting that large changes consistently introduce higher integration complexity.

These findings indicate that integration difficulty is associated with the size of AI-generated changes. Larger PRs are more likely to interfere with concurrent development activity, leading to a higher probability of textual merge conflicts. This finding highlights the importance of controlling change size in AI-assisted development workflows to reduce integration friction.

\begin{resultbox}
\noindent\textbf{\faLightbulbO \hspace{0.01in} Key takeaway: }
Integration difficulty increases with the size of AI-generated changes, as larger PRs are more prone to merge conflicts.
\end{resultbox}

\section{Related Work}

\nd \textbf{Pull-Based Development and Code Review.} Pull-based development has become the dominant contribution model in open source ecosystems. Gousios et al.~\cite{Gousios:2014} conducted one of the first large-scale empirical studies of the pull-based model, analyzing review practices, acceptance rates, and integration dynamics. Later work examined review quality, reviewer behavior, and factors influencing pull request acceptance~\cite{kononenko2016code,11024268,11025917,men:2025:codereviewing}. While these studies provide valuable insight into collaborative workflows, they typically do not reconstruct merge outcomes at the commit level. As a result, integration friction due to textual conflicts is not explicitly captured in most pull request datasets.

\textsc{AgenticFlict} complements prior PR research by introducing conflict-aware metadata that can be integrated with review and acceptance analyses, filling a critical gap in understanding how modern automated and agentic contributions impact repository health~\cite{watanabe2025useagenticcodingempirical}.

\nd \textbf{AI-Assisted and AI-Generated Code Contributions.} The emergence of large language models for code generation has motivated empirical research on AI-assisted programming. Controlled experiments show that developers complete tasks faster when assisted by systems such as GitHub Copilot~\cite{peng2023impact}. Human-computer interaction studies examine developer expectations and usability challenges of code generation tools~\cite{vaithilingam2022expectation}. More recently, large-scale mining studies have begun to analyze repositories containing AI-generated or AI-assisted contributions~\cite{ogenrwot2026aicodingagentsmodify,watanabe2025useagenticcodingempirical,li2025aidev,horikawa2025agentic}. These studies examine acceptance rates, code quality, and maintenance characteristics. However, they do not explicitly study merge outcomes or quantify textual conflict severity. 

Our work provides a large-scale dataset of reproducible textual merge conflict labels and fine-grained conflict-region metadata for Agentic PRs.

\nd \textbf{Merge Conflicts in Collaborative Development}.

Merge conflicts have long been recognized as a significant source of coordination overhead in distributed software development~\cite{mahmood:2020:elastic,brun2013early}. The impact of merge conflicts extends beyond individual repositories. In large software ecosystems, structural interconnectedness means that integration friction in one component can propagate widely~\cite{10.1007/978-3-032-08649-5_9}, reinforcing the need for systematic study of conflict-introducing contributions. \citet{brun2013early} demonstrate that collaboration conflicts are frequent and costly, and propose early detection mechanisms to mitigate their impact. Subsequent studies have analyzed the causes and characteristics of merge conflicts in large-scale repositories, highlighting the role of concurrent edits, file centrality, and developer coordination patterns~\cite{10.1145/3377811.3380344,10356713}. Research has also investigated conflict prediction and prevention techniques~\cite{10.1145/3377811.3380344}. These approaches leverage historical commit data, code ownership, and file modification patterns to estimate the likelihood of conflicts prior to merging. However, existing conflict datasets primarily focus on human-authored changes and do not explicitly consider contributions generated by AI coding agents, despite recent evidence that AI assistants can increase commit frequency by approximately 13.55\%~\cite{cui2026effects}.

\textsc{AgenticFlict} extends this line of research by providing a reproducible dataset of textual merge conflicts specifically in the context of Agentic PRS.

\nd \textbf{Datasets and Benchmarks.}
Existing merge conflict datasets can be broadly categorized into traditional collaborative benchmarks and emerging AI-centric repositories. Traditional benchmarks focus on human-authored conflicts at scale, such as the 2,731 Java-based projects studied by \citet{8468085}, reporting that nearly 20\% of merges require manual intervention and subsequent analyses of conflict structure in 123 Java Projects, revealing that conflicts are primarily concentrated within shared method bodies~\cite{Accioly:2018:semi}. 
More recently, ConflictBench~\cite{SHEN2024112084} was introduced as a dedicated benchmark specifically designed to evaluate merge tools. It provides a curated collection of conflicting scenarios, categorized by programming language and conflict type. Similarly, datasets like those used in SBCR~\cite{10.1145/3748256} focus on the textual similarity between conflict resolutions and their parent versions, offering nearly 10,000 conflict chunks for 1,062 Java projects.

With the rise of Large Language Models (LLMs), AI-centric datasets have emerged to capture interactions between developers and LLMs. DevGPT~\cite{DevGPT:2024} introduced a dataset of shared ChatGPT conversations linked to GitHub artifacts, later extended by PatchTrack~\cite{ogenrwot2025patchtrackcomprehensiveanalysischatgpts,ogenrwot:2024ase} with additional PRs to study the influence of ChatGPT on pull request decision-making and developer-ChatGPT conversation lifecycle. Similarly, AIDev~\cite{li2025aidev} provides a large-scale collection of contributions from multiple AI coding agents. While these datasets offer valuable insights into how AI-generated contributions are created and reviewed, they primarily focus on high-level metadata such as acceptance rates and discussion dynamics. They do not provide explicit or reproducible labels for textual merge conflicts, limiting our ability to systematically quantify the integration friction introduced by AI-generated changes.

\section{Threats to Validity}

In this section, we discuss potential threats to the validity of the \textsc{AgenticFlict} dataset.

First, conflict detection is based on deterministic local merge simulation using commit identifiers retrieved via the GitHub GraphQL API. In some cases, these references may no longer be available due to repository changes such as force pushes or deletions. We exclude such PRs to avoid incorrect conflict labeling, at the cost of reduced coverage. 

Second, we capture merge conflicts using textual conflict markers produced by Git during merge simulation. While this provides a consistent and widely used proxy for integration issues, it does not account for higher-level forms of conflict such as logical inconsistencies or post-merge defects. Furthermore, our analysis focuses on open and closed (unmerged) PRs, which means we may miss conflicts that were previously encountered and resolved during the lifecycle of merged PRs. We adopt this design for a methodological reason: once a PR is merged, Git no longer preserves the pre-merge conflict state. Any conflicts that arose during review were resolved prior to merging, meaning that the conflict markers, line spans, and severity metrics we extract cannot be recovered post-hoc from the merge commit alone. Reconstructing them would require speculative replay of intermediate review states, which we deliberately avoid to preserve reproducibility

Finally, \textsc{AgenticFlict} is constructed on top of the AIDev dataset and therefore inherits its limitations. In particular, the dataset focuses on Agentic PRS and may overrepresent repositories that actively adopt AI tools. As a result, our findings may not generalize to all open-source projects or industrial settings. Extending the dataset to include merge conflicts from human-authored PRs is an important direction for future work. In addition, conflict behavior may vary across programming languages, repository sizes, and development practices.

\section{Conclusion}
In this paper, we introduced \textsc{AgenticFlict}, a large-scale dataset designed to characterize merge conflicts in AI coding agent PRs. The dataset comprises over 142K Agentic PRs collected from more than 59K repositories, with over 107K successfully analyzed through deterministic merge simulation resulting in over 29K (27.67\%) PRs exhibiting textual merge conflicts. Our approach enables reproducible conflict detection and provides fine-grained conflict-region metadata, including conflicting files and line-level spans, resulting in over 336K conflict regions. Our analysis shows that merge conflicts are both frequent and often substantial in AI-generated contributions, highlighting integration as a key challenge in AI-assisted software development. By making these conflict patterns observable at scale, \textsc{AgenticFlict} provides a foundation for studying how AI agents interact with collaborative development workflows. In future work, we plan to extend the dataset to include merge conflicts from human-authored PRs in the same repositories as the Agentic PRs, enabling direct comparative analysis. We plan to investigate conflict characteristics that go beyond textual markers, including logical inconsistencies and post-merge defects, which our current pipeline does not capture. We hope this dataset will support future research on conflict prediction, automated resolution, and the design of tools that better integrate AI-generated code into modern development pipelines. More broadly, our work contributes to understanding the evolving role of AI agents as active participants in Software Engineering 3.0.

\begin{acks}
This research was supported by the National Science Foundation Grant\#: 2519136 and the National Science Foundation Major Research Instrumentation (NSF MRI) (Grant\#: 2117941).
\end{acks}
\bibliographystyle{ACM-Reference-Format}
\bibliography{references}

@String{Computing = "Computing" }

@String{Computer = "{IEEE} Computer" }

@String{Springer = "Springer-Verlag" }

@inproceedings{vaithilingam2022expectation,
  title        = {Expectation vs. Experience: Evaluating the Usability of Code Generation Tools Powered by Large Language Models},
  author       = {Vaithilingam, Priyan and Zhang, Tianyi and Glassman, Elena L.},
  booktitle    = {Proceedings of the 2022 CHI Conference on Human Factors in Computing Systems},
  pages        = {1--17},
  year         = {2022},
  publisher    = {ACM},
address = {USA}
}

@inproceedings{vaithilingam2023copilot,
  title        = {Copilot or Co-Author? Examining the Role of Code Generation Tools in Collaborative Programming},
  author       = {Vaithilingam, Priyan and Xu, Zheng and Glassman, Elena L.},
  booktitle    = {Proceedings of the 2023 IEEE Symposium on Visual Languages and Human-Centric Computing (VL/HCC)},
  year         = {2023},
  publisher    = {IEEE},
address = {USA}
}

@article{li2025aidev,
title={{The Rise of AI Teammates in Software Engineering (SE) 3.0: How Autonomous Coding Agents Are Reshaping Software Engineering}}, 
author={Li, Hao and Zhang, Haoxiang and Hassan, Ahmed E.},
journal={arXiv preprint arXiv:2507.15003},
year={2025}
}

@article{horikawa2025agentic,
  title={Agentic Refactoring: An Empirical Study of AI Coding Agents},
  author={Horikawa, Kosei and Li, Hao and Kashiwa, Yutaro and Adams, Bram and Iida, Hajimu and Hassan, Ahmed E},
  journal={arXiv preprint arXiv:2511.04824},
  year={2025}
}

@misc{githubCopilot,
  author       = {{GitHub Copilot}},
  title        = {GitHub Copilot},
  howpublished = {\url{https://github.com/copilot}},
  note         = {Accessed: 2025-12-14},
  year         = {2025}
}

@misc{openaiCodexWeb,
  author       = {{OpenAI}},
  title        = {Codex — OpenAI},
  howpublished = {\url{https://openai.com/codex/}},
  note         = {Accessed: 2025-12-14},
  year         = {2025}
}

@misc{devinAIWeb,
  author       = {{Devin AI}},
  title        = {Devin AI — AI Coding Assistant},
  howpublished = {\url{https://app.devin.ai/}},
  note         = {Accessed: 2025-12-14},
  year         = {2025}
}

@misc{cursor2025,
  author       = {{Cursor}},
  title        = {Cursor: AI Code Editor},
  howpublished = {\url{https://cursor.com/}},
  year         = {2025},
  note         = {Accessed: 2025-12-14}
}

@misc{claudeAI2025,
  author       = {{Anthropic}},
  title        = {Claude.ai},
  howpublished = {\url{https://claude.ai/}},
  year         = {2025},
  note         = {Accessed: 2025-12-14}
}

@inproceedings{ogenrwot:2024ase,
author = {Ogenrwot, Daniel and Businge, John},
title = {PatchTrack: Analyzing ChatGPT's Impact on Software Patch Decision-Making in Pull Requests},
year = {2024},
isbn = {9798400712487},
publisher = {Association for Computing Machinery},
address = {New York, NY, USA},
url = {https://doi.org/10.1145/3691620.3695338},
doi = {10.1145/3691620.3695338},
booktitle = {Proceedings of the 39th IEEE/ACM International Conference on Automated Software Engineering},
pages = {2480–2481},
numpages = {2},
location = {Sacramento, CA, USA},
series = {ASE '24}
}

@misc{ogenrwot2025patchtrackcomprehensiveanalysischatgpts,
      title={PatchTrack: A Comprehensive Analysis of ChatGPT's Influence on Pull Request Outcomes}, 
      author={Daniel Ogenrwot and John Businge},
      year={2025},
      eprint={2505.07700},
      archivePrefix={arXiv},
      primaryClass={cs.SE},
      url={https://arxiv.org/abs/2505.07700}, 
}

@misc{hassan2025agenticsoftwareengineeringfoundational,
      title={Agentic Software Engineering: Foundational Pillars and a Research Roadmap}, 
      author={Ahmed E. Hassan and Hao Li and Dayi Lin and Bram Adams and Tse-Hsun Chen and Yutaro Kashiwa and Dong Qiu},
      year={2025},
      eprint={2509.06216},
      archivePrefix={arXiv},
      primaryClass={cs.SE},
      url={https://arxiv.org/abs/2509.06216}, 
}

@misc{hassan2024ainativesoftwareengineeringse,
      title={Towards AI-Native Software Engineering (SE 3.0): A Vision and a Challenge Roadmap}, 
      author={Ahmed E. Hassan and Gustavo A. Oliva and Dayi Lin and Boyuan Chen and Zhen Ming and Jiang},
      year={2024},
      eprint={2410.06107},
      archivePrefix={arXiv},
      primaryClass={cs.SE},
      url={https://arxiv.org/abs/2410.06107}, 
}

@misc{watanabe2025useagenticcodingempirical,
      title={On the Use of Agentic Coding: An Empirical Study of Pull Requests on GitHub}, 
      author={Miku Watanabe and Hao Li and Yutaro Kashiwa and Brittany Reid and Hajimu Iida and Ahmed E. Hassan},
      year={2025},
      eprint={2509.14745},
      archivePrefix={arXiv},
      primaryClass={cs.SE},
      url={https://arxiv.org/abs/2509.14745}, 
}

@article{businge:emse:2022,
	author = {Businge, John and Openja, Moses and Nadi, Sarah and Berger, Thorsten},
	doi = {10.1007/s10664-021-10078-2},
	journal = {Journal of Empirical Software Engineering},
	title = {Reuse and Maintenance Practices among Divergent Forks in Three Software Ecosystems},
	volume = {27},
	number = {2},
	pages = {54},
	year = {2022}
}

@inproceedings{businge:blockchain:2022,
	author = {Henrique Rocha and John Businge},
	booktitle = {5th International Workshop on Blockchain Oriented Software Engineering},
	title = {Blockchain-Oriented Software Variant Forks: A Preliminary Study},
	year = {2022}}

@inproceedings{Businge:benevol:2020,
  author    = {John Businge and
               Alexandre Decan and
               Ahmed Zerouali and
               Tom Mens and
               Serge Demeyer},
  editor    = {Mike Papadakis and
               Maxime Cordy},
  title     = {An Empirical Investigation of Forks as Variants in the npm Package
               Distribution},
  booktitle = {Proceedings of the 19th Belgium-Netherlands Software Evolution Workshop,
               {BENEVOL} 2020, Luxembourg, December 3-4, 2020},
  series    = {{CEUR} Workshop Proceedings},
  volume    = {2912},
  publisher = {CEUR-WS.org},
  year      = {2020},
  url       = {http://ceur-ws.org/Vol-2912/paper1.pdf},
  bibsource = {dblp computer science bibliography, https://dblp.org}
}

@inproceedings{businge:2018icsme,
	author = {John Businge and Moses Openja and Sarah Nadi and Engineer Bainomugisha and Thorsten Berger},
	booktitle = {International Conference on Software Maintenance and Evolution},
	pages = {625--634},
	publisher = {IEEE},
	title = {Clone-Based Variability Management in the {{Android}} Ecosystem},
	year = {2018}}

@ARTICLE{brun2013early,
  author={Brun, Yuriy and Holmes, Reid and Ernst, Michael D. and Notkin, David},
  journal={IEEE Transactions on Software Engineering}, 
  title={Early Detection of Collaboration Conflicts and Risks}, 
  year={2013},
  volume={39},
  number={10},
  pages={1358-1375},
  doi={10.1109/TSE.2013.28}}

@inproceedings{Kononenko2016code,
author = {Kononenko, Oleksii and Baysal, Olga and Godfrey, Michael W.},
title = {Code review quality: how developers see it},
year = {2016},
isbn = {9781450339001},
publisher = {Association for Computing Machinery},
address = {New York, NY, USA},
url = {https://doi.org/10.1145/2884781.2884840},
doi = {10.1145/2884781.2884840},
booktitle = {Proceedings of the 38th International Conference on Software Engineering},
pages = {1028–1038},
numpages = {11},
location = {Austin, Texas},
series = {ICSE '16}
}

@inproceedings{Gousios:2014,
author = {Gousios, Georgios and Pinzger, Martin and Deursen, Arie van},
title = {An exploratory study of the pull-based software development model},
year = {2014},
isbn = {9781450327565},
publisher = {Association for Computing Machinery},
address = {New York, NY, USA},
url = {https://doi.org/10.1145/2568225.2568260},
doi = {10.1145/2568225.2568260},
booktitle = {Proceedings of the 36th International Conference on Software Engineering},
pages = {345–355},
numpages = {11},
location = {Hyderabad, India},
series = {ICSE 2014}
}

@article{peng2023impact,
  title={The Impact of AI on Developer Productivity: Evidence from GitHub Copilot},
  author={Peng, Sida and Kalliamvakou, Eirini and Cihon, Peter and Demirer, Mert},
  journal={arXiv preprint arXiv:2302.06590},
  year={2023},
  note={Submitted on 13 Feb 2023},
  url={https://doi.org/10.48550/arXiv.2302.06590},
  doi={10.48550/arXiv.2302.06590}
}

@inproceedings{Vijayvergiya2024AI,
author = {Vijayvergiya, Manushree and Salawa, Ma\l{}gorzata and Budiseli\'{c}, Ivan and Zheng, Dan and Lamblin, Pascal and Ivankovi\'{c}, Marko and Carin, Juanjo and Lewko, Mateusz and Andonov, Jovan and Petrovi\'{c}, Goran and Tarlow, Daniel and Maniatis, Petros and Just, Ren\'{e}},
title = {AI-Assisted Assessment of Coding Practices in Modern Code Review},
year = {2024},
isbn = {9798400706851},
publisher = {Association for Computing Machinery},
address = {New York, NY, USA},
url = {https://doi.org/10.1145/3664646.3665664},
doi = {10.1145/3664646.3665664},
booktitle = {Proceedings of the 1st ACM International Conference on AI-Powered Software},
pages = {85–93},
numpages = {9},
location = {Porto de Galinhas, Brazil},
series = {AIware 2024}
}

@INPROCEEDINGS{6227180,
  author={Guimarães, Mário Luís and Silva, António Rito},
  booktitle={2012 34th International Conference on Software Engineering (ICSE)}, 
  title={Improving early detection of software merge conflicts}, 
  year={2012},
  volume={},
  number={},
  pages={342-352},
  doi={10.1109/ICSE.2012.6227180}}

@ARTICLE{9625780,
  author={Vale, Gustavo and Hunsen, Claus and Figueiredo, Eduardo and Apel, Sven},
  journal={IEEE Transactions on Software Engineering}, 
  title={Challenges of Resolving Merge Conflicts: A Mining and Survey Study}, 
  year={2022},
  volume={48},
  number={12},
  pages={4964-4985},
  doi={10.1109/TSE.2021.3130098}}

@INPROCEEDINGS{8094445,
  author={McKee, Shane and Nelson, Nicholas and Sarma, Anita and Dig, Danny},
  booktitle={2017 IEEE International Conference on Software Maintenance and Evolution (ICSME)}, 
  title={Software Practitioner Perspectives on Merge Conflicts and Resolutions}, 
  year={2017},
  volume={},
  number={},
  pages={467-478},
  doi={10.1109/ICSME.2017.53}}

@INPROCEEDINGS{8870173,
  author={Owhadi-Kareshk, Moein and Nadi, Sarah and Rubin, Julia},
  booktitle={2019 ACM/IEEE International Symposium on Empirical Software Engineering and Measurement (ESEM)}, 
  title={Predicting Merge Conflicts in Collaborative Software Development}, 
  year={2019},
  volume={},
  number={},
  pages={1-11},
  doi={10.1109/ESEM.2019.8870173}}

@article{Shen2023A,
author = {Shen, Bowen and Gulzar, Muhammad Ali and He, Fei and Meng, Na},
title = {A Characterization Study of Merge Conflicts in Java Projects},
year = {2023},
issue_date = {March 2023},
publisher = {Association for Computing Machinery},
address = {New York, NY, USA},
volume = {32},
number = {2},
issn = {1049-331X},
url = {https://doi.org/10.1145/3546944},
doi = {10.1145/3546944},
journal = {ACM Trans. Softw. Eng. Methodol.},
month = mar,
articleno = {40},
numpages = {28}
}

@misc{ogenrwot2026aicodingagentsmodify,
      title={How AI Coding Agents Modify Code: A Large-Scale Study of GitHub Pull Requests}, 
      author={Daniel Ogenrwot and John Businge},
      year={2026},
      eprint={2601.17581},
      archivePrefix={arXiv},
      primaryClass={cs.SE},
      url={https://arxiv.org/abs/2601.17581}, 
}

@article{SHEN2024112084,
title = {ConflictBench: A benchmark to evaluate software merge tools},
journal = {Journal of Systems and Software},
volume = {214},
pages = {112084},
year = {2024},
issn = {0164-1212},
doi = {https://doi.org/10.1016/j.jss.2024.112084},
url = {https://www.sciencedirect.com/science/article/pii/S0164121224001298},
author = {Bowen Shen and Na Meng}
}

@ARTICLE{8468085,
  author={Ghiotto, Gleiph and Murta, Leonardo and Barros, Márcio and van der Hoek, André},
  journal={IEEE Transactions on Software Engineering}, 
  title={On the Nature of Merge Conflicts: A Study of 2,731 Open Source Java Projects Hosted by GitHub}, 
  year={2020},
  volume={46},
  number={8},
  pages={892-915},
  doi={10.1109/TSE.2018.2871083}}

@inproceedings{10.1145/3540250.3549163,
author = {Svyatkovskiy, Alexey and Fakhoury, Sarah and Ghorbani, Negar and Mytkowicz, Todd and Dinella, Elizabeth and Bird, Christian and Jang, Jinu and Sundaresan, Neel and Lahiri, Shuvendu K.},
title = {Program merge conflict resolution via neural transformers},
year = {2022},
isbn = {9781450394130},
publisher = {Association for Computing Machinery},
address = {New York, NY, USA},
url = {https://doi.org/10.1145/3540250.3549163},
doi = {10.1145/3540250.3549163},
booktitle = {Proceedings of the 30th ACM Joint European Software Engineering Conference and Symposium on the Foundations of Software Engineering},
pages = {822–833},
numpages = {12},
location = {Singapore, Singapore},
series = {ESEC/FSE 2022}
}

@INPROCEEDINGS{ogenrwot:2025:scam,
  author={Ogenrwot, Daniel and Businge, John},
  booktitle={2025 IEEE International Conference on Source Code Analysis \& Manipulation (SCAM)}, 
  title={Refactoring-Aware Patch Integration Across Structurally Divergent Java Forks}, 
  year={2025},
  volume={},
  number={},
  pages={25-36},
  doi={10.1109/SCAM67354.2025.00010}}

@inproceedings{di2017software,
  title={Software heritage: Why and how to preserve software source code},
  author={Di Cosmo, Roberto and Zacchiroli, Stefano},
  booktitle={iPRES 2017},
  year={2017}
}

@INPROCEEDINGS{GHTorrent,
  author={Gousios, Georgios},
  booktitle={2013 10th Working Conference on Mining Software Repositories (MSR)}, 
  title={The GHTorent dataset and tool suite}, 
  year={2013},
  volume={},
  number={},
  pages={233-236},
  doi={10.1109/MSR.2013.6624034}}

@INPROCEEDINGS{Svajlenko:2014:ICSME,
  author={Svajlenko, Jeffrey and Islam, Judith F. and Keivanloo, Iman and Roy, Chanchal K. and Mia, Mohammad Mamun},
  booktitle={2014 IEEE International Conference on Software Maintenance and Evolution}, 
  title={Towards a Big Data Curated Benchmark of Inter-project Code Clones}, 
  year={2014},
  volume={},
  number={},
  pages={476-480},
  doi={10.1109/ICSME.2014.77}}

@misc{github-graphql-doc,
  author       = {{GitHub, Inc.}},
  title        = {GraphQL API},
  year         = {2026},
  howpublished = {\url{https://docs.github.com/en/graphql}},
  note         = {Accessed: 2026-04-02}
}

@misc{github-rate-limit,
  author       = {{GitHub, Inc.}},
  title        = {Rate limits and query limits for the GraphQL API},
  year         = {2026},
  howpublished = {https://docs.github.com/en/graphql/overview/rate-limits-and-query-limits-for-the-graphql-api},
  note         = {Accessed: 2026-04-02}
}

@misc{github-tos,
  author       = {{GitHub, Inc.}},
  title        = {GitHub Terms of Service},
  year         = {2026},
  howpublished = {https://docs.github.com/en/site-policy/github-terms/github-terms-of-service},
  note         = {Accessed: 2026-04-02}
}

@misc{git-partial-clone,
  author       = {{Git Project}},
  title        = {Partial Clone},
  year         = {2026},
  howpublished = {https://git-scm.com/docs/partial-clone},
  note         = {Accessed: 2026-04-02}
}

@inproceedings{mahmood:2020:elastic,
author = {Mahmood, Wardah and Chagama, Moses and Berger, Thorsten and Hebig, Regina},
title = {Causes of merge conflicts: a case study of ElasticSearch},
year = {2020},
isbn = {9781450375016},
publisher = {Association for Computing Machinery},
address = {New York, NY, USA},
url = {https://doi.org/10.1145/3377024.3377047},
doi = {10.1145/3377024.3377047},
booktitle = {Proceedings of the 14th International Working Conference on Variability Modelling of Software-Intensive Systems},
articleno = {9},
numpages = {9},
location = {Magdeburg, Germany},
series = {VaMoS '20}
}

@article{cui2026effects,
  title={The effects of generative AI on high-skilled work: Evidence from three field experiments with software developers},
  author={Cui, Kevin Zheyuan and Demirer, Mert and Jaffe, Sonia and Musolff, Leon and Peng, Sida and Salz, Tobias},
  journal={Management Science},
  year={2026},
  publisher={INFORMS}
}

@inproceedings{10.1145/3377811.3380344,
author = {Brindescu, Caius and Ahmed, Iftekhar and Leano, Rafael and Sarma, Anita},
title = {Planning for untangling: predicting the difficulty of merge conflicts},
year = {2020},
isbn = {9781450371216},
publisher = {Association for Computing Machinery},
address = {New York, NY, USA},
url = {https://doi.org/10.1145/3377811.3380344},
doi = {10.1145/3377811.3380344},
booktitle = {Proceedings of the ACM/IEEE 42nd International Conference on Software Engineering},
pages = {801–811},
numpages = {11},
location = {Seoul, South Korea},
series = {ICSE '20}
}

@INPROCEEDINGS{10356713,
  author={Vale, Gustavo and Fernandes, Eduardo and Figueiredo, Eduardo and Apel, Sven},
  booktitle={2023 IEEE 23rd International Working Conference on Source Code Analysis and Manipulation (SCAM)}, 
  title={Behind Developer Contributions on Conflicting Merge Scenarios}, 
  year={2023},
  volume={},
  number={},
  pages={25-36},
  doi={10.1109/SCAM59687.2023.00014}}

@inproceedings{10.1145/3555228.3555229,
author = {Campos Junior, Heleno de S. and de Menezes, Gleiph Ghiotto L. and Barros, M\'{a}rcio de Oliveira and van der Hoek, Andr\'{e} and Murta, Leonardo Gresta Paulino},
title = {Towards Merge Conflict Resolution by Combining Existing Lines of Code},
year = {2022},
isbn = {9781450397353},
publisher = {Association for Computing Machinery},
address = {New York, NY, USA},
url = {https://doi.org/10.1145/3555228.3555229},
doi = {10.1145/3555228.3555229},
booktitle = {Proceedings of the XXXVI Brazilian Symposium on Software Engineering},
pages = {425–434},
numpages = {10},
location = {Virtual Event, Brazil},
series = {SBES '22}
}

@article{Accioly:2018:semi,
	author = {Accioly, Paola and Borba, Paulo and Cavalcanti, Guilherme},
	date = {2018/08/01},
	doi = {10.1007/s10664-017-9586-1},
	id = {Accioly2018},
	isbn = {1573-7616},
	journal = {Empirical Software Engineering},
	number = {4},
	pages = {2051--2085},
	title = {Understanding semi-structured merge conflict characteristics in open-source Java projects},
	url = {https://doi.org/10.1007/s10664-017-9586-1},
	volume = {23},
	year = {2018}}

@inproceedings{DevGPT:2024,
author = {Xiao, Tao and Treude, Christoph and Hata, Hideaki and Matsumoto, Kenichi},
title = {DevGPT: Studying Developer-ChatGPT Conversations},
year = {2024},
isbn = {9798400705878},
publisher = {Association for Computing Machinery},
address = {New York, NY, USA},
url = {https://doi.org/10.1145/3643991.3648400},
doi = {10.1145/3643991.3648400},
booktitle = {Proceedings of the 21st International Conference on Mining Software Repositories},
pages = {227–230},
numpages = {4},
location = {Lisbon, Portugal},
series = {MSR '24}
}

@article{10.1145/3748256,
author = {Campos Junior, Heleno de S. and Ghiotto L. de Menezes, Gleiph and Barros, M\'{a}rcio de Oliveira and van der Hoek, Andr\'{e} and Murta, Leonardo Gresta Paulino},
title = {Towards a feasible evaluation function for search-based merge conflict resolution},
year = {2025},
publisher = {Association for Computing Machinery},
address = {New York, NY, USA},
issn = {1049-331X},
url = {https://doi.org/10.1145/3748256},
doi = {10.1145/3748256},
note = {Just Accepted},
journal = {ACM Trans. Softw. Eng. Methodol.},
month = jul
}

@INPROCEEDINGS{11024268,
  author={Alami, Adam and Ernst, Neil},
  booktitle={2025 IEEE/ACM 18th International Conference on Cooperative and Human Aspects of Software Engineering (CHASE)}, 
  title={Human and Machine: How Software Engineers Perceive and Engage with AI-Assisted Code Reviews Compared to Their Peers}, 
  year={2025},
  volume={},
  number={},
  pages={63-74},
  doi={10.1109/CHASE66643.2025.00016}}

@INPROCEEDINGS{11025917,
  author={Gonçalves, Pavlína Wurzel and Rani, Pooja and Storey, Margaret-Anne and Spinellis, Diomidis and Bacchelli, Alberto},
  booktitle={2025 IEEE/ACM 33rd International Conference on Program Comprehension (ICPC)}, 
  title={Code Review Comprehension: Reviewing Strategies Seen Through Code Comprehension Theories}, 
  year={2025},
  volume={},
  number={},
  pages={589-601},
  doi={10.1109/ICPC66645.2025.00068}}

@article{men:2025:codereviewing,
	author = {G{\"o}{\c c}men, Ismail Sergen and Cezayir, Ahmed Salih and T{\"u}z{\"u}n, Eray},
	date = {2025/02/12},
	doi = {10.1007/s10664-024-10600-2},
	id = {G{\"o}{\c c}men2025},
	isbn = {1573-7616},
	journal = {Empirical Software Engineering},
	number = {3},
	pages = {64},
	title = {Enhanced code reviews using pull request based change impact analysis},
	url = {https://doi.org/10.1007/s10664-024-10600-2},
	volume = {30},
	year = {2025}}

@inproceedings{10.1145/2597073.2597074,
author = {Kalliamvakou, Eirini and Gousios, Georgios and Blincoe, Kelly and Singer, Leif and German, Daniel M. and Damian, Daniela},
title = {The promises and perils of mining GitHub},
year = {2014},
isbn = {9781450328630},
publisher = {Association for Computing Machinery},
address = {New York, NY, USA},
url = {https://doi.org/10.1145/2597073.2597074},
doi = {10.1145/2597073.2597074},
booktitle = {Proceedings of the 11th Working Conference on Mining Software Repositories},
pages = {92–101},
numpages = {10},
location = {Hyderabad, India},
series = {MSR 2014}
}

@INPROCEEDINGS{5069475,
  author={Bird, Christian and Rigby, Peter C. and Barr, Earl T. and Hamilton, David J. and German, Daniel M. and Devanbu, Prem},
  booktitle={2009 6th IEEE International Working Conference on Mining Software Repositories}, 
  title={The promises and perils of mining git}, 
  year={2009},
  volume={},
  number={},
  pages={1-10},
  doi={10.1109/MSR.2009.5069475}}

@dataset{AgenticFlict2026Replication,
  author       = {Ogenrwot, Daniel and
                  Businge, John},
  title        = {AgenticFlict: A Large-Scale Dataset of Merge
                   Conflicts in AI Coding Agent Pull Requests on
                   GitHub
                  },
  month        = may,
  year         = 2026,
  publisher    = {Zenodo},
  doi          = {10.5281/zenodo.19396916},
  url          = {https://doi.org/10.5281/zenodo.19396916},
}

@inproceedings{10.1145/3520312.3534864,
author = {Ziegler, Albert and Kalliamvakou, Eirini and Li, X. Alice and Rice, Andrew and Rifkin, Devon and Simister, Shawn and Sittampalam, Ganesh and Aftandilian, Edward},
title = {Productivity assessment of neural code completion},
year = {2022},
isbn = {9781450392730},
publisher = {Association for Computing Machinery},
address = {New York, NY, USA},
url = {https://doi.org/10.1145/3520312.3534864},
doi = {10.1145/3520312.3534864},
booktitle = {Proceedings of the 6th ACM SIGPLAN International Symposium on Machine Programming},
pages = {21–29},
numpages = {9},
location = {San Diego, CA, USA},
series = {MAPS 2022}
}

@ARTICLE{7887704,
  author={Cosentino, Valerio and Cánovas Izquierdo, Javier L. and Cabot, Jordi},
  journal={IEEE Access}, 
  title={A Systematic Mapping Study of Software Development With GitHub}, 
  year={2017},
  volume={5},
  number={},
  pages={7173-7192},
  doi={10.1109/ACCESS.2017.2682323}}

@inproceedings{10.1145/3540250.3549112,
author = {Ramkisoen, Poedjadevie Kadjel and Businge, John and van Bladel, Brent and Decan, Alexandre and Demeyer, Serge and De Roover, Coen and Khomh, Foutse},
title = {PaReco: patched clones and missed patches among the divergent variants of a software family},
year = {2022},
isbn = {9781450394130},
publisher = {Association for Computing Machinery},
address = {New York, NY, USA},
url = {https://doi.org/10.1145/3540250.3549112},
doi = {10.1145/3540250.3549112},
booktitle = {Proceedings of the 30th ACM Joint European Software Engineering Conference and Symposium on the Foundations of Software Engineering},
pages = {646–658},
numpages = {13},
location = {Singapore, Singapore},
series = {ESEC/FSE 2022}
}

@Inbook{Businge:chap:2023,
author="Businge, John
and Abdi, Mehrdad
and Demeyer, Serge",
editor="Mens, Tom
and De Roover, Coen
and Cleve, Anthony",
title="Analyzing Variant Forks of Software Repositories from Social Coding Platforms",
bookTitle="Software Ecosystems: Tooling and Analytics",
year="2023",
publisher="Springer International Publishing",
address="Cham",
pages="131--152",
isbn="978-3-031-36060-2",
doi="10.1007/978-3-031-36060-2_6",
url="https://doi.org/10.1007/978-3-031-36060-2_6"
}

@InProceedings{10.1007/978-3-032-08649-5_9,
author="Ogenrwot, Daniel
and Businge, John
and Arifuzzaman, Shaikh",
editor="Rahimi, Nick
and Margapuri, Venkat
and Golilarz, Noor Amiri",
title="Structural and Connectivity Patterns in the Maven Central Software Dependency Network",
booktitle="Software and Data Engineering",
year="2026",
publisher="Springer Nature Switzerland",
address="Cham",
pages="129--151",
isbn="978-3-032-08649-5"
}

\end{document}